\documentclass[11pt]{article}
\usepackage[T1]{fontenc}
\usepackage{amsfonts,amsmath,amsthm,amssymb,dsfont,mathtools}
\usepackage[margin=1in]{geometry}
\usepackage{fullpage}
\usepackage[colorlinks,citecolor=blue,linkcolor=blue,urlcolor=red,pagebackref]{hyperref}
\usepackage[capitalise]{cleveref}
\usepackage{url}
\usepackage{enumerate}
\usepackage[linesnumbered,boxed,ruled,vlined]{algorithm2e}
\usepackage{thm-restate}
\usepackage{graphicx}
\usepackage{mathdots}
\usepackage{paralist}
\usepackage[normalem]{ulem}
\usepackage{xspace}
\xspaceaddexceptions{]\}}
\usepackage{tabu}
\usepackage{framed}
\usepackage{float,wrapfig}
\usepackage{subfigure}
\usepackage{tikz}
\usepackage[usenames,dvipsnames]{pstricks}

\newtheorem{mytheorems}{Theorem}[section]
\newtheorem{them}[mytheorems]{Theorem}
\newtheorem{lem}[mytheorems]{Lemma}
\newtheorem{prop}[mytheorems]{Proposition}
\newtheorem{cor}[mytheorems]{Corollary}
\newtheorem{claim}[mytheorems]{Claim}
\theoremstyle{Definition}
\newtheorem{definition}[mytheorems]{Definition}

\newtheorem{remark}[mytheorems]{Remark}
\newtheorem{obs}[mytheorems]{Observation}

\renewcommand{\epsilon}{\varepsilon}
\newcommand{\eps}{\varepsilon}
\renewcommand{\Pr}{\operatorname*{\mathbf{Pr}}}

\newcommand{\poly}{\operatorname{\mathrm{poly}}}
\newcommand{\polylog}{\poly\log}

\newcommand{\R}{\mathbb{R}}
\newcommand{\N}{\mathbb{N}}

\newcommand{\Z}{\mathbb{Z}}
\newcommand{\opt}{\mathrm{OPT}}
\newcommand{\sol}{\mathrm{SOL}}
\newcommand{\subs}{Subset Sum\xspace}
\newcommand{\parti}{Partition\xspace}
\newcommand{\caS}{\mathcal{S}\xspace}

\title{Approximating Knapsack and Partition via Dense Subset Sums}

\author{Mingyang Deng \thanks{dengm@mit.edu} \\ MIT \and Ce Jin \thanks{cejin@mit.edu. Partially supported by NSF Grant CCF-2129139} \\ MIT\and Xiao Mao \thanks{matthew99a@gmail.com}\\ MIT}

\begin{document}

	\setcounter{page}{0} \clearpage
	\maketitle
	\thispagestyle{empty}

	\begin{abstract}
	 \emph{Knapsack} and \emph{Partition} are two important additive problems whose fine-grained complexities in the $(1-\eps)$-approximation setting are not yet settled. In this work, we make progress on both problems by giving improved algorithms.
	 \begin{itemize}
	     \item \emph{Knapsack} can be $(1 - \eps)$-approximated in $\tilde O(n  + (1/\epsilon) ^ {2.2} )$ time, 
	     improving the previous $\tilde O(n  + (1/\epsilon) ^ {2.25} )$  by Jin (ICALP'19).
	     There is a known conditional lower bound of $(n+1/\eps)^{2-o(1)}$ based on $(\min,+)$-convolution hypothesis.
	     \item \emph{Partition} can be $(1 - \eps)$-approximated in $\tilde O(n  + (1/\epsilon) ^ {1.25} )$ time,
	     improving the previous $\tilde O(n  + (1/\epsilon) ^ {1.5} )$  by Bringmann and Nakos (SODA'21).
	     There is a known conditional lower bound of $(1/\eps)^{1-o(1)}$ based on Strong Exponential Time Hypothesis.
	 \end{itemize}
	 Both of our new algorithms apply the additive combinatorial results on dense subset sums by Galil and Margalit (SICOMP'91), Bringmann and Wellnitz (SODA'21). Such techniques have not been explored in the context of Knapsack prior to our work. In addition, we design several new methods to speed up the divide-and-conquer steps which naturally arise in solving additive problems.
	\end{abstract}

	\newpage

	\section{Introduction}

    \subsection{Background}
    \emph{Knapsack}, \emph{Subset Sum}, and \emph{Partition} are three fundamental problems in
    computer science and mathematical optimization, and are actively being studied in fields such as  integer programming and fine-grained complexity. 
In the Knapsack problem (sometimes also called 0-1 Knapsack), we are given a set $I$ of $n$ items where each item $i \in I$ has weight $w_i>0$ and profit $p_i>0$, as well as a knapsack capacity $W$, and we want to choose a subset $J\subseteq I$ satisfying the weight constraint $\sum_{j\in J}w_j \le W$ such that the total profit $\sum_{j\in J}p_j$ is maximized.
The Subset Sum problem is a special case of Knapsack, where the weight of an item is always equal to their profit. The Partition problem is a special case of Subset Sum, where the capacity equals half of the total weight of the items. In other words, in \parti we want to partition the input items into two parts so that their sums is as close as possible.  

These three problems are well-known to be hard: they appeared in Karp's original list of 21 NP-hard problems \cite{karp1972reducibility}.
To cope with NP-hardness, a natural direction is to study their \emph{approximation algorithms}. Given a parameter $\eps > 0$, and an input instance with optimal value $\opt$, a $(1-\eps)$-approximation algorithm is required to output a number $\sol$ such that $(1 - \eps)\opt \le \sol \le \opt$.
Fortunately, these three problems are well-known to have \emph{fully polynomial-time approximation schemes (FPTASes)}, namely $(1-\eps)$-approximation algorithm that runs in $\poly(n,1/\eps)$ time, for any $\eps>0$. 

There has been a long line of research since the 70's on getting approximation schemes for these problems with  improved time complexities in terms of $n$ and $1/\eps$  \cite{ibarra1975fast, lawler1979fast, DBLP:conf/mfcs/GensL79,  kellerer1999new,DBLP:journals/jcss/KellererMPS03, kellerer2004improved, rhee2015faster,jansen2018faster, chan2018approximation,MuchaW019, DBLP:conf/icalp/Jin19,BN21,DBLP:conf/icalp/BringmannC22}. On  the other hand, recent advances in fine-grained complexity have pointed out the limit of such improvements, under well-believed hardness assumptions \cite{cygan2019problems, kunnemann2017fine,  abboud2019seth,BN21}. 
Here, we briefly describe the most recent results along this line.
\begin{itemize}
    \item \textbf{Knapsack:}  The best known algorithm by Jin \cite{DBLP:conf/icalp/Jin19} runs in $\tilde O(n+\eps^{-2.25})$ time, and is based on the previous algorithm of Chan \cite{chan2018approximation} in $\tilde O(n+\eps^{-2.4})$ time.   \cite{cygan2019problems} and   \cite{kunnemann2017fine} showed a conditional lower bound of  $(n+\frac{1}{\eps})^{2-o(1)}$, based on   the $(\min,+)$-convolution hypothesis.
    
    \item \textbf{Subset Sum:} The best known algorithm by Bringmann and Nakos \cite{BN21} runs in $\tilde O(n+\eps^{-2}/2^{\Omega(\sqrt{\log(1/\eps)})})$ time (improving \cite{DBLP:journals/jcss/KellererMPS03} by low-order factors). Bringmann and Nakos \cite{BN21} showed a matching lower bound based on the $(\min,+)$-convolution hypothesis.
    
    \item \textbf{Partition:} The first breakthrough by   Mucha, W\k{e}grzycki and W\l{}odarczyk \cite{MuchaW019} gave a randomized algorithm in  
$\tilde O(n+1/\eps^{5/3})$ time. Later, Bringmann and Nakos \cite{BN21} improved it to deterministic $\tilde O(n+1/\eps^{-3/2})$ time. 
 Abboud, Bringmann, Hermelin, and Shabtay \cite{abboud2019seth} showed that \parti cannot be approximated in $\poly(n)/\eps^{1-\delta}$ time for any $\delta>0$, under the Strong Exponential Time Hypothesis. 
   \end{itemize}

We can see that the complexity of Subset Sum is already settled, but for Knapsack and Partition there still remain gaps between the best known algorithms and their conditional lower bounds.

	\subsection{Our Results}
	In this work, we make progress on this direction, by giving improved approximation schemes for Knapsack and Partition.
        \begin{them} \label{theo:main}
 There is a randomized algorithm for  $(1 - \eps$)-approximating Knapsack with running time \footnote{Throughout this paper,
we use $\tilde O(f)$ to denote $O(f\cdot \text{poly} \log (f))$. } \[\tilde O\left (n  + \epsilon ^ {-11/5} / 2 ^ {\Omega(\sqrt{\log(1 /  \epsilon)})}\right ).\] 
        \end{them}

        	\begin{them} \label{theo:parti}
    There is a deterministic algorithm for $(1-\eps)$-approximating \parti  with running time \[\tilde O\left (n + \eps^{-5/4}\right ).\] 
	\end{them}
        
    \subsection{Technical Overview}

	A useful result in additive combinatorics for many subset sum related problems is the one from by Galil and Margalit \cite{DBLP:journals/siamcomp/GalilM91}, which was later improved by Bringmann and Wellnitz \cite{bw21} (based on results of S{\'a}rk{\"o}zy \cite{sarkozy1994fine}). These combinatorial results reveal structural properties on the set $\mathcal{S}(X)$ of subset sums of a positive integer set $X$, defined as $\mathcal{S}(X):= \{\sum_{y\in Y}y : Y \subseteq X\}$, in the case when $X$ is \emph{``dense''}. Intuitively, it states that if the total number of items is large, then a large portion of the subset sum can be computed very efficiently, so only a small margin of the sumsets with value up to some $\lambda$ needs to be approximated. We apply these combinatorial results to Partition and, surprisingly, to the ``two-dimensional'' problem of Knapsack where each item has both weights and values.

    \subsubsection{Knapsack}
    \paragraph*{Faster knapsack via dense subset sums}
    Our improved approximation scheme for knapsack relies on multiple technical components from the previous algorithms by Chan \cite{chan2018approximation} and Jin \cite{DBLP:conf/icalp/Jin19}. However, one novel key idea that makes our improvement possible is a technique previously not explored in the context of knapsack algorithms: the additive combinatorics result for dense subset sum by Galil and Margalit \cite{DBLP:journals/siamcomp/GalilM91}. One particular result useful to us roughly says the following: when $X$ consists of $n$ distinct integers in $[\ell,2\ell]$ for a small enough $\ell \ll n^2$, then there is a long enough interval $[\lambda, (\sum_{x\in X}x) - \lambda]$ that is \emph{densely filled} with elements in $\mathcal{S}(X)$, in the sense that every two adjacent elements must be very close to each other. A formal version of the statement is in \cref{lemma:density}. 
     As we will see, such density statements will be useful in the framework of Jin \cite{DBLP:conf/icalp/Jin19}. Jin's approximation algorithm for Knapsack separately deals with items with high and low \emph{efficiency}, defined as the profit-to-weight ratio $p_i/w_i$. Intuitively, it is not very profitable to include too many low-efficiency items in the solution. Indeed, after some technical steps, Jin managed to place an upper bound $B$ on the total profit contributed by low-efficiency items in any optimal solution, so that it is still correct to only compute the answers for low-efficiency items up to $B$ (which would take much shorter time than original).  The way Jin proved such a bound $B$ was by a certain greedy exchange argument: given a solution set with too many low-efficiency items (which occupies a total capacity of $W_L$), remove them and try to fill in the freed up space of $W_L$ using high-efficiency items instead. This would potentially lead to a better solution, contradicting the optimality of the given solution.  
    
    Naturally, such an exchange is not always profitable, since the high-efficiency items may not be able to fill up the entire space $W_L$. Jin's argument accounts for this issue by additionally making sure that all items have size in an interval $[\ell,2\ell]$, so that the wasted space after the exchange cannot be larger $2\ell$ (otherwise one can always fit in another high-efficiency item). In our new algorithm, we refine this argument using combinatorial results on dense subset sums: observe that the task of minimizing the wasted space is equivalent to a subset sum problem on the sizes of high-efficiency items.  By setting up parameters appropriately, we can make sure that the dense subset sum result applies, leading to a much smaller wasted space.    
    
    Having refined this argument, we can improve Jin \cite{DBLP:conf/icalp/Jin19} by putting a stricter upper bound on the total profit contributed by low-efficiency items, leading to an improved running time.

    \subsubsection{Partition}
	\paragraph*{Densified divide and conquer}  
	There have been many FPTAS algorithms for problems such as \subs, \parti, \textsc{Knapsack} that employ the technique of divide and conquer, e.g.\ \cite{chan2018approximation,  DBLP:conf/icalp/Jin19,BN21}. Unlike previous methods, our improvement crucially relies on performing divide-and-conquer on the \emph{sorted} list of items. To motivate our idea, we note that in most divide and conquer based algorithms, the bottleneck to the running time is incurred at the bottom levels, where we need to merge two sets of answers often with the same complexity of those at the top levels. However, intuitively, if our list is sorted, at the bottom levels the items have values contained in a small interval, and hence the sumsets of these items are clustered in several small intervals with large gaps between them.
	To exploit this property, instead of using the usual 1D FFT to merge the sumsets, we ``densify'' the sumsets and merge them using 2D FFT, so that our running time is only dependent on the total length of these small intervals. 
	
	We note that the idea of 2D-FFT has been employed before to \subs by Koiliaris and Xu \cite{kx19}, but it is used in a different spirit: while we use 2D-FFT to ``densify'' sumsets, in \cite{kx19} it is used to bound the size of the solution for each sum.\footnote{In fact, it can be verified that by combining our way of doing 2D-FFT with the other techniques in \cite{kx19} we can get an alternative deterministic solution for \subs running in $\tilde O(\sqrt{n}t)$ time. It is interesting to see if the two ways of using 2D-FFT can be combined to improve the running time for \subs deterministically.}
	\paragraph*{Combining with additive combinatorics result}  
	The additive combinatorics result for dense subset sum by Galil and Margalit \cite{DBLP:journals/siamcomp/GalilM91} has also been used to an extent in the $\tilde O(n + 1 / \epsilon ^ {5 / 3})$ algorithm by \cite{MuchaW019}. In our algorithm we combine this with densified divide and conquer. Note that $\lambda$ can be much smaller than the total sum $\sigma$ of the items, and this would mean that an $\eps\sigma$ additive error is an $\eps' \lambda$ additive error for a much larger $\eps' = \eps \sigma/\lambda$, so we only need to ensure $(1-\eps')$-multiplicative approximation instead of $(1-\eps)$ during the computation.

        \subsection{Paper Organization}
        We will give useful definitions and lemmas in \cref{sec:prelim}. In \cref{sec:knapsack} we present our algorithm for Knapsack.
        In \cref{sec:partition} we present our algorithm for Partition. Some standard reductions and known lemmas from previous works are deferred to appendix.

    \section{Preliminaries}
    \label{sec:prelim}

We write $\N = \{0,1,2,\dots\}$ and $\N^+ = \{1,2,\dots\}$. For $n\in \N$ we write $[n] = \{1,2,\dots,n\}$.

\subsection{Problem Statements}

    In the Knapsack problem, the input is a list of $n$ items $(p_1,w_1),\dots,(p_n,w_n) \in \N\times \N$ together with a knapsack capacity $W \in \N$, and the optimal value is 
    \[\opt:= \max_{J\subseteq [n]} \Big \{\sum_{j\in J }p_j \, \Big \vert \,  \sum_{j\in J}w_j \le W \Big \}.\]

	In the easier Partition problem,  the input is a list  of $n$ integers $x_1,\dots,x_n \in \N$, and the optimal value is 
    \[\opt:= \max_{J\subseteq [n]} \Big \{\sum_{j\in J }x_j  \,\Big \vert \,  \sum_{j\in J}x_j \le \frac{1}{2}\sum_{i\in [n]}x_i \Big \}.\]

Given a Knapsack (or a Partition) instance and a parameter $\eps \in (0,1)$, an \emph{$(1-\eps)$-approximation algorithm} is required to output a number $\sol$ such that $(1 - \eps)\opt \le \sol \le \opt$.

In both problems, we can assume $n =O(\eps^{-4})$ and hence $\log n = O(\log \eps^{-1})$.  For larger $n$,  Lawler's  algorithm \cite{lawler1979fast} for Knapsack in $O(n\log \frac{1}{\eps} + (\frac{1}{\eps})^4)$ time is already near-optimal. 

We will sometimes describe algorithms with approximation ratio $1-O(\eps)$ (or $1 - \eps \cdot \polylog(1/\eps)$), which can be made $1-\eps$ by scaling down $\eps$ by a constant factor (or a logarithmic factor) at the beginning. 

\subsection{Sumsets and Subset Sums}

In a multiset $A$, an element $a$ could appear multiple times (the number of times it appears is the \emph{multiplicity} of $a$ in $A$). We use $A\uplus B$ to denote union without removing duplicates (i.e., possibly resulting in a multiset). 
	
For a multiset $Y\subset \N$,  let $\Sigma(Y)=\sum_{y\in Y} y$ denote the sum of its elements (without removing duplicates).
	
For a multiset $X\subset \N$, let $\caS(X) = \{\Sigma(Y) : Y\subseteq X\} $ be the set of its subset sums, and let $\caS(X;t) = \caS(X) \cap [0,t]$ be the set of its subset sums up to $t$.

For a number $c$ and a set $X$, define $c\cdot X= \{ cx: x\in X\}$. For two sets $X,Y$, define their \emph{sumset} $X+Y = \{x+y : x\in X, y\in Y\}$. Given sets $X\subseteq [n],Y\subseteq[n]$, the sumset $X+Y$ can be computed in $O(n \log n)$ time using FFT. This simple fact has a straightforward generalization to 2 dimension, which we state below.
\begin{lem}[2-dimensional FFT, e.g., {\cite[Chapter 12.8]{blahut2010fast}}]
\label{lem:2dfft}
Given two sets $A_1,A_2 \subseteq [n] \times [m]$, one can compute
\[A_1+A_2:= \big \{ (x_1+x_2,y_1+y_2) : (x_1,y_1)\in A_1,(x_2,y_2)\in A_2\big  \}\]
in $ O(nm \log(nm))$ time deterministically.
\end{lem}

\subsection{Knapsack Problem and Profit functions}
\label{sec:prelimknap}
In the knapsack problem,  assume $0<w_i\le W$ and $p_i>0$ for every item $i$. Then a trivial lower bound of the maximum total profit is $\max_j p_j$. At the beginning, we can discard all items $i$ with $p_i \le \frac{\eps}{n} \max_j p_j$, reducing the total profit by at most $\eps \max_j p_j$, which is only an $O(\eps)$ fraction of the optimal total profit. So we can assume $\frac{\max_j p_j}{\min_j p_j} \le \frac{n}{\eps}$.

For a set $I$ of items, we use $f_I$ to denote its \emph{profit function}, defined as
\begin{equation*}
f_I(x) = \max \Bigg \{ \sum_{i\in J} p_i  : \sum_{i \in J} w_i\le x, \;\; J \subseteq I \Bigg\}
\end{equation*}
over $x\in [0,+\infty)$. Note that $f_I$ is a monotone nondecreasing step function. Adopting the terminology of Chan \cite{chan2018approximation}, the \textit{complexity} of a monotone step function refers to the number of its steps.

Let $I_1,I_2$ be two disjoint subsets of items, and $I = I_1 \uplus I_2$. It is straightforward to see that $f_{I} = f_{I_1} \oplus f_{I_2}$, where $\oplus$ denotes $(\max,+)$-convolution, defined by 
$(f\oplus g)(x) = \max_{0\le x'\le x}(f(x')+g(x-x'))$.

    \subsection{$(1-\delta,\Delta)$ approximation up to $t$}
    Both our algorithms for Knapsack and Partition frequently use the notion of \emph{$(1-\delta,\Delta)$-approximation up to $t$}. Their definitions are analogous, as stated below.

    \begin{definition}[Approximation for Profit Functions]
          For functions $\tilde f, f$ and real numbers $t,\Delta \in \R_{\ge 0}, \delta \in [0,1)$, we say that  $\tilde f$ is a  \emph{$(1-\delta,\Delta)$ approximation of $f$ up to $t$}, if 
          \[ \tilde f(w)\le f(w)\]
          holds for all $w\ge 0$, and 
          \[ \tilde f(w) \ge (1-\delta)f(w) - \Delta\]
          holds whenever $f(w)\le t, w\ge 0$.
    \end{definition} 

The following notion of approximation will be useful in our \parti algorithm. Similar notions have been termed as ``weak approximation'' in the literature \cite{MuchaW019, BN21}, in contrast to ``strong approximation'' that would be required for approximating general \subs instances. 
\begin{definition}[Approximation for Integer Sets]
For integer sets $A, B \subseteq \N$, and real numbers $t,\Delta \in \R_{\ge 0}, \delta \in [0,1)$, we say that $A$ is a \emph{$(1-\delta,\Delta)$ approximation of $B$ up to $t$}, if \begin{enumerate}
    \item  for every $b\in B \cap [0,t]$, there exists $a\in A$ such that $(1-\delta)b-\Delta \le a\le b$, and,
    \item for every $a\in A $, there exists $b\in B$ such that $(1-\delta)b-\Delta\le a \le b$.
\end{enumerate}   One can assume $A \subseteq \N \cap [0,t]$ in this case without loss of generality.
\end{definition}

For the case of $t=+\infty$, we simply omit the phrase ``up to $t$''. 

We also refer to $(1,\Delta)$ approximation as \emph{$\Delta$-additive approximation}, and refer to $(1-\delta,0)$ approximation as \emph{$(1-\delta)$-multiplicative approximation}, or simply \emph{$(1-\delta)$ approximation}.

We have the following simple facts regarding approximating merged sumsets and profit functions.
\begin{prop}
\label{prop:merge}
For $i\in \{1,2\}$, suppose $A_i$ is a $(1-\delta,\Delta_i)$ approximation of $\caS(X_i)$ up to $t$.
Then, $(A_1+A_2) \cap [0,t]$ is a $(1-\delta,\Delta_1+\Delta_2)$ approximation of $\caS(X_1\uplus X_2)$ up to $t$.
\end{prop}
\begin{proof}
For any $b\in \caS(X_1\uplus X_2) \cap [0,t]$ where $b=b_1+b_2$ for $b_i\in \caS(X_i) \cap [0,t]$ ($i\in \{1,2\}$), there exits $a_i \in A_i$ such that $(1-\delta)b_i - \Delta_i \le a_i \le b_i$. Hence, $a_1+a_2\le b_1+b_2=b \le t$, and $a_1+a_2 \ge (1-\delta)b_1 - \Delta_1+(1-\delta)b_2 - \Delta_2  = (1-\delta)b - (\Delta_1+\Delta_2)$.

The converse direction can be verified similarly.
\end{proof}

The following fact can be proved similarly.
\begin{prop}
\label{prop:merge2}
For $i\in \{1,2\}$, suppose $\tilde f_i$ is a $(1-\delta,\Delta_i)$ approximation of the profit function $f_{I_i}$ up to $t$.
Then, $(\tilde f_1 \oplus \tilde f_2)$ is a $(1-\delta,\Delta_1+\Delta_2)$ approximation of $f_{I_1\uplus I_2}$ up to $t$.
\end{prop}

Following Chan \cite{chan2018approximation} and Jin \cite{DBLP:conf/icalp/Jin19}, given a monotone step function $f$ (we sometimes also call it a profit function, although it might not be equal to the profit function $f_I$ of any particular item set $I$)  with range contained in $\{0\} \cup [A,B]$, one can
round $f$ down to powers of $1/(1-\eps)$, and obtain another profit function $\tilde f$ which has complexity only $O(\eps^{-1}\log (B/A))$, and $(1-\eps)$-approximates $f$. 
In our algorithm we will always have $B/A \le \poly(n/\eps)$,  so we may always assume that the intermediate profit functions computed during our algorithm are  monotone step functions with complexity $\tilde O(\eps^{-1})$, by incurring $(1-\eps)$ approximation factor each time.

\subsection{Additive Combinatorics}
We need several results on dense subset sums developed by a series of works including \cite{sarkozy1994fine,DBLP:journals/siamcomp/GalilM91,lev2003blocks,bw21}.
        The following structural lemma follows from Theorem 4.1 and Theorem 4.2 of Bringmann and Wellnitz \cite{bw21}.
        \begin{restatable}{lem}{densityrestate}
            Let $n$ distinct positive integers $X=\{x_1,\dots,x_n\} \subseteq [\ell,2\ell]$ be given, where $\ell = o(n^2 /\log n)$.
            
           Then, for a universal constant $c\ge 1$, for every $c\ell^2/n \le t \le \Sigma(X)/2$, there exists $t ^ {\prime} \in \caS(X)$ such that $0\le t^\prime - t  \le 8\ell / n$.  
            \label{lemma:density}
        \end{restatable}
        A proof of \cref{lemma:density} is included in \cref{app:density}.

The following algorithmic lemma follows from the main theorem of \cite{bw21}, and will be used in our Partition algorithm.
\begin{lem}[Follows from \cite{bw21}]
Given $n$ distinct positive integers $X=\{x_1,\dots,x_n\} \subseteq [\ell,2\ell]$, there exists $\lambda = \tilde \Theta  (\ell ^ 2/n ) $ such that, if $\lambda \le \Sigma(X)/2$, then in $\tilde O(n)$ time we can construct a deterministic data structure supporting the following query in $O(1)$ time: given $L,R$ such that $\lambda \le L \le R \le \Sigma(X)/2$, report whether there exists $t\in [L,R]$ such that $t \in \caS(X)$.  
\label{theo:galil}
\end{lem}
\begin{remark}
We remark that the main theorem stated in \cite{bw21} only supports querying whether a given $\lambda \le t \le \Sigma(X)/2$ is a subset sum.  In our application, we require a version supporting range queries. This is easy to achieve by building an additional prefix sum array in the proof of \cite[Theorem 4.6]{bw21}, which supports range sum queries.
\end{remark}

    \section{Approximating Knapsack}
    
    \label{sec:knapsack}
    \subsection{Known Lemmas}
    By known reductions (e.g., \cite{chan2018approximation,DBLP:conf/icalp/Jin19}), we can focus on solving the following cleaner problem, which already captures the main difficulty of knapsack.
    
\begin{restatable}{prob}{proknaprestate}
\label{probknap}
Assume $\eps\in (0,1/2)$ and $1/\eps \in \N^+$. Given a list $I$ of items $(p_1,w_1),\dots, (p_n,w_n)$ with weights $w_i\in \N$ and profits $p_i$ being multiples of $\eps$ in the interval $[1,2)$, compute a profit function that  $(1-\eps)$-approximates $f_I$ up to $2/\eps$. 
\end{restatable}

\begin{restatable}{lem}{redknap}
	\label{lem:reduction-knap} If for some $c\ge 2$, \cref{probknap} can be solved in $\tilde O(n + 1/\eps^c)$ time, then $(1-\eps)$-approximating \emph{Knapsack} can also  be solved in $\tilde O(n + 1/\eps^c)$ time. 
\end{restatable}
\cref{lem:reduction-knap} will be proved in the appendix.

 Based on Chan's techniques \cite{chan2018approximation}, Jin \cite{DBLP:conf/icalp/Jin19} obtained the following lemmas for $(1-\eps)$-approximating knapsack up to a small $B$ or when there are few distinct values $p_i$.
 
        \begin{lem} [Follows from Lemma 17 of \cite{DBLP:conf/icalp/Jin19}] \label{theo:jin1}
            Given a list $I$ of items $(p_1,w_1),\dots, (p_n,w_n)$ with weights $w_i\in \N$ and profits $p_i$ being multiples of $\eps$ in the interval $[1,2)$, one can $(1-\eps)$-approximate the profit function $f_I$ up to $B$ in $\tilde O(n+\epsilon ^ {-2}B ^ {1 / 3} / 2 ^ {\Omega(\sqrt{\log(1 /  \epsilon)})})$ time.        \end{lem}
        
        \begin{lem} [Follows from Theorem 3 of {\cite{chan2018approximation}}] \label{theo:jin2}
        Given a list $I$ of items $(p_1,w_1),\dots, (p_n,w_n)$ with weights $w_i\in \N$ and profits $p_i$ being multiples of $\eps$ in the interval $[1,2)$, if there are only $m$ distinct profit values $p_i$,  then one can $(1-\eps)$-approximate the profit function $f_I$ in $\tilde O(n+\epsilon ^ {-3 / 2}m  / 2 ^ {\Omega(\sqrt{\log(1 /  \epsilon)})})$ time.
\footnote{In the proceedings version of our paper, we incorrectly claimed that the task in \cref{theo:jin2} can be done in $\tilde O(n+\epsilon ^ {-3 / 2}m^{3/4}  / 2 ^ {\Omega(\sqrt{\log(1 /  \epsilon)})})$ time. Here, the statement of \cref{theo:jin2} has been corrected.  As a result, several parameters in \cref{sec:greedyproof} have been adjusted accordingly.
 This correction did not affect the final time bound of our main result (\cref{theo:main}), since the step that invokes  \cref{theo:jin2} is not a bottleneck in our algorithm.}
        \end{lem}

        The following useful lemma allows us to merge multiple profit functions, which was proved by Chan using divide-and-conquer and improved algorithms for $(\min,+)$-convolution \cite{bremner2014necklaces,williams2014faster,chan2016deterministic}.
\begin{lem}[\text{\cite[Lemma 2(i)]{chan2018approximation}}]
\label{dc}
Let $f_1,\dots, f_m$ be monotone step functions with total complexity $O(n)$ and
ranges contained in $\{ 0\} \cup [A, B]$. Then we can compute a monotone step function that has complexity
$\tilde O(\frac{1}{\eps}\log B/A)$ and 
$(1-O(\eps))$-approximates $f_1 \oplus\dots \oplus f_m$, in $O(n) + \tilde O((\frac{1}{\eps})^2 m/2^{\Omega(\sqrt{\log(1/\eps)})} \log B/A)$ time.
\end{lem}

        \subsection{Greedy Exchange Argument via Dense Subset Sum}
        \label{sec:greedyproof}

        The goal of this section is to prove the following \cref{lem:greedy}. Our algorithm is based on a greedy exchange argument similar to \cite[Lemma 20]{DBLP:conf/icalp/Jin19}, but we can obtain better bounds by combining with number theoretic results on dense subset sums.
        \begin{lem}
        \label{lem:greedy}
        Given a list $I$ of $n$ items with $p_i$ being multiples of $\eps$ in interval $[1,2)$, and integer $1\le m \le n$ with $m=O(1/\eps)$, one can compute in $O(n+\eps^{-11/5}/2^{\Omega(\sqrt{\log (1/\eps)})} )$ time a profit function that $(m\eps)$-additively approximates $f_I$ up to $2m$.
        \end{lem}
        
        The proof of \cref{lem:greedy} assumes the following ingredient, which will be proved in later sections using random partitioning.
        \begin{restatable}{lem}{knapsackmainrestate}
         \label{lemma:knapmain}
         Given a list $I$ of $n= O(1/\eps)$ items with $p_i$ being multiples of $\eps$ in interval $[1,2)$, one can compute in $\tilde O(n ^ {{4}/{5}}\epsilon ^ {-{7}/{5}}/2^{\Omega(\sqrt{\log (1/\eps)})})$ time a profit function that $(n\eps)$-additively approximates $f_I$.
        \end{restatable}

Now we proceed to describe the algorithm for \cref{lem:greedy}. Given items $(p_1,w_1),\dots,(p_n,w_n)$, where $p_i\in[1,2)$ are multiples of $\eps$, we sort them by non-increasing order of efficiency, $p_1/w_1\ge p_2/w_2\ge\dots \ge p_n/w_n$. 
        Then, we consider prefixes of this sequence of items, and define the following measure of diversity:
        \begin{definition}[$D(i)$]
           For $1\le i \le n$, let $D(i)= \min_{J} C([i]\setminus J)$, where the minimization is over all subsets $J \subseteq [i]$ with $|J|\le 2m$, and $C([i] \setminus J)$ denote the number of distinct values in $\{p_j : j\in [i] \setminus J\}$. 
        \end{definition}
        We have the following immediate observations about $D(i)$.
        \begin{obs}
            \begin{enumerate}
                \item For all $2\le i\le n$, $0\le D(i) - D(i-1)\le 1$.
                \item $D(i)$ (and the minimizer $J$) can be computed in $\tilde O(i)$ time by the following greedy algorithm:  Start with all values $p_1,p_2,\dots,p_i$.  Repeat the following up to $2m$ times: remove the value $p_j$ with the minimum multiplicity, and add $j$ into $J$. 
            \end{enumerate}
            \label{obs:D}
        \end{obs}

        Now, we set parameter $ \Delta = \lfloor\epsilon ^ {-5 / 8}\rfloor $. Define $i \in \{1,2,\dots,n\}$ to be the maximum such that $D(i) \le \Delta$, which can be found using \cref{obs:D} with a binary search in $\tilde O(n)$ time. 

The following lemma is the key component in our proof of \cref{lem:greedy}. 
            \begin{lem}[Greedy Exchange Lemma]
            \label{lem:exchange}
                 Let $S\subseteq [n]$ 
                 be any item set with total profit $\sum_{s\in S}p_s \le 2m$. 
                 Let $B:=9c\eps^{-1}/\Delta$, where $c\ge 1$ is the universal constant in \cref{lemma:density}.
                 
                 Then, there exists an item set $\tilde S \subseteq [n]$,  such that the total profit $\tilde p$ contributed by items $[n] \setminus [i]$ in $\tilde S$ satisfies
                 \begin{equation}
                 \label{eqn:req1}
                 \tilde  p:=\sum_{s \in \tilde S \cap ([n] \setminus [i])} p_s \le B,
                 \end{equation}
                       and
                       \begin{equation}
                 \label{eqn:req2}
                      \sum_{s \in \tilde  S } p_s  \ge (1-\eps) \sum_{s \in   S } p_s,
                       \end{equation}
                      and
                      \begin{equation}
                 \label{eqn:req3}
                      \sum_{s \in \tilde  S } w_s  \le  \sum_{s \in   S } w_s.
                      \end{equation}
            \end{lem}
            \begin{proof}
                If $D(i) <\Delta$, then by the definition of $i$ we have $i=n$, and we can simply let $\tilde S=S$, since $\tilde p=0$ always holds. So in the following we assume $D(i)=\Delta$.

                We define $\tilde S\subseteq [n]$ as the maximizer of 
                \[ \sum_{s\in \tilde S \cap [i]}p_s + \sum_{s\in \tilde S \cap ([n] \setminus [i])}(1-\eps)p_s\]
                among all $\tilde S$ satisfying $\sum_{s\in \tilde S} w_s \le \sum_{s\in S} w_s$ and $\sum_{s\in \tilde S} p_s \le \sum_{s\in S} p_s$.
                 We claim that $\tilde S$ satisfies the properties \eqref{eqn:req1}, \eqref{eqn:req2}, \eqref{eqn:req3}. Observe that \eqref{eqn:req2}, \eqref{eqn:req3} immediately follow from the definition of $\tilde S$. The main part is to verify \eqref{eqn:req1}.

Suppose for contradiction that \eqref{eqn:req1} does not hold. Then, we can find a subset $K\subseteq \tilde S \cap ([n]\setminus [i])$ with total profit $p^{*} = \sum_{k\in K} p_k \in (B,B+2]$, which can  be obtained by removing items from $\tilde S\cap ([n]\setminus [i])$ (recall that each item has profit in $[1,2)$).

        Define item set $I ^ {\prime} := [i]\setminus \tilde S$.
         Since $|\tilde S| < \sum_{s\in \tilde S}p_s/\min_{s\in \tilde S} p_s \le \sum_{s\in \tilde S}p_s \le \sum_{s\in  S}p_s \le 2m$, 
         by the definition of $D(i)$, we know that $\{p_i: i\in I ^ {\prime}\}$ contains at least $D(i)=\Delta$ distinct elements.

         We apply \cref{lemma:density} on the set of integers $X=\{p_i/\eps : i\in I^\prime \} \subseteq [1/\eps,2/\eps)$ which contains at least $\Delta$ distinct integers, where the premise $1/\eps = o(\Delta^2/\log \Delta)$ in \cref{lemma:density} is satisfied by our choice of $\Delta = \lfloor \eps^{-5/8}\rfloor$. \cref{lemma:density} states that for every $t\in [c\eps^{-2}/\Delta, 0.5\Delta/\eps]$, there exists $t'\in \caS(X)$ such that $0\le t'-t\le 8\eps^{-1}/\Delta$. Here we set \[t:=\frac{(1-\eps) p^{*}}{\eps} + \frac{\eps^{-1}}{\Delta},\] which satisfies $t>p^{*}(1-\eps)/\eps > (1-\eps)B/\eps  = (1-\eps)(9c\eps^{-1}/\Delta)/\eps > c\eps^{-2}/\Delta$, and $t < (B+2)/\eps + \eps^{-1}/\Delta = 9c\eps^{-2}/\Delta + 2/\eps +\eps^{-1}/\Delta\le O(\eps^{-11/8}) \le 0.5\Delta/\eps$. Then the conclusion of \cref{lemma:density} says that there is a subset  $R\subseteq I'$ of items with total profit $\tilde p:= \eps \cdot t'$, satisfying 
         \begin{equation}
            1/\Delta\le \tilde p - p^{*}(1-\eps)\le 9/\Delta. \label{eqn:my}
         \end{equation}
         Note that \eqref{eqn:my} implies
         \begin{align*}
             p^{*} - \tilde p &\ge \eps\cdot p^{*} - 9/\Delta\\
             & > \eps \cdot B -9/\Delta\\
             &  = \eps \cdot 9c \eps^{-1}/\Delta - 9/\Delta\\
             & \ge 0.
         \end{align*}

         Recall that
          $R\subseteq I' = [i]\setminus \tilde S$
           and $K\subseteq \tilde S \cap ([n]\setminus [i])$,
            which must both be non-empty.
          Since the efficiency of items are sorted in non-increasing order, we have $\min_{r\in R}p_r/w_r \ge \max_{k\in K} p_k/w_k$. Now we define the set of items \[\tilde S':= (\tilde S \setminus K) \cup R. \]
         Then, we have
         \begin{align*}
            \sum_{s\in  \tilde S}p_s - \sum_{s\in \tilde S'}p_s &= \sum_{k\in  K}p_k - \sum_{r\in R}p_r \\
            & = p^* -\tilde p\\
            & \ge 0,
         \end{align*}
         and
         \begin{align*}
            \sum_{s\in  \tilde S}w_s - \sum_{s\in \tilde S'}w_s &= \sum_{k\in  K}w_k - \sum_{r\in R}w_r \\
            & \ge \frac{\sum_{k\in K}p_k}{\max_{k\in K}(p_k/w_k)} - \frac{\sum_{r\in R}p_r}{\min_{r\in R}(p_r/w_r)}\\
            & \ge  \frac{1}{\min_{r\in R}(p_r/w_r)}\cdot \left ( \sum_{k\in K}p_k - \sum_{r\in R}p_r\right )\\
 &  = \frac{1}{\min_{r\in R}(p_r/w_r)}\cdot \left ( p^{*} - \tilde p\right )\\
 & \ge 0.
         \end{align*}
         Hence, $\sum_{s\in \tilde S'}p_s \le \sum_{s\in \tilde S}p_s$ and $\sum_{s\in \tilde S'}w_s \le \sum_{s\in \tilde S}w_s$.
          On the other hand, by \eqref{eqn:my}, we know that
         \begin{align*}
 &                \left (\sum_{s\in \tilde S' \cap [i]}p_s + \sum_{s\in \tilde S' \cap ([n] \setminus [i])}(1-\eps)p_s\right ) - \left (\sum_{s\in \tilde S \cap [i]}p_s + \sum_{s\in \tilde S \cap ([n] \setminus [i])}(1-\eps)p_s\right )\\
 = \ &  \sum_{r\in R} p_r - \sum_{k\in K}(1-\eps) p_k\\
 = \ & \tilde p - (1-\eps)p^*\\
 \ge \ & 1/\Delta > 0,
 \end{align*}
 contradicting the definition of $\tilde S$ being a  maximizer.
 
 Hence, we have established that $\tilde S$ satisfies \eqref{eqn:req1}.
            \end{proof}

        Now we are ready to prove \cref{lem:greedy}.
        \begin{proof}[Proof of \cref{lem:greedy}]
        Recall that $i \in \{1,2,\dots,n\}$ is the maximum such that $D(i) \le \Delta$, which can be found using \cref{obs:D} with a binary search in $\tilde O(n)$ time.  Let $J \subset[i]$ with $|J|\le 2m$ be the minimizer for $D(i)$.
        
        Now, we approximately compute the profit functions $f_{J}, f_{[i]\setminus J}, f_{[n]\setminus [i]}$ for three item sets $J, [i]\setminus J, [n]\setminus [i]$ using different algorithms, described as follows:
        \begin{enumerate}
            \item Use \cref{lemma:knapmain} to compute $f_1$ that $(2m\eps)$-additively approximates $f_J$, in $O(m ^ {\frac{4}{5}}\epsilon ^ {-\frac{7}{5}} / 2 ^ {\Omega(\sqrt{\log(1 /  \epsilon)})}) \le O(\epsilon ^ {-\frac{11}{5}} / 2 ^ {\Omega(\sqrt{\log(1 /  \epsilon)})})$ time.
            \item By definition of $i$, items in $[i]\setminus J$ have no more than $\Delta$ distinct profit values. Hence we can use \cref{theo:jin2} to compute $f_2$ that $(1-\eps)$-approximates  $f_{[i]\setminus J}$, in $\tilde O(\Delta \epsilon ^ {-3 / 2}) = \tilde O(\epsilon ^ {-17 / 8})$ time.
            \item Use \cref{theo:jin1} to compute $f_3$ that $(1-\eps)$-approximates the $f_{[n]\setminus [i]}$ up to $B=\Theta(\eps^{-1}/\Delta)$ (defined in \cref{lem:exchange}), in $\tilde O(B^{1/3}\eps^{-2}) \le \tilde O(\epsilon ^ {-17 / 8})$ time.
        \end{enumerate}
            Finally, merge the three parts $f_1,f_2,f_3$ using \cref{dc} in $\tilde O(\eps ^ {-2})$ time, and return the result.\footnote{Although the running time of the second and third algorithm is dominated by the first algorithm, a simple rebalancing of parameters does not seem to yield better complexity, due to various constraints in the parameter settings for \cref{lemma:knapmain}. }
            
        In the third part, the correctness of only computing up to $B$ is justified by \cref{lem:exchange}, which shows that if we only consider approximating sets with total profit up to $2m$, then we can assume the items in $[n] \setminus [i]$ only contributes profit at most $B$ \eqref{eqn:req1}, at the cost of only incurring an $(1-\eps)$ approximation factor \eqref{eqn:req2}.
        
        To analyze the error, notice that in the first part we incur an additive error of $(2m\eps)$. In the second and third part and the final merging step we incur $(1-O(\eps))$ multiplicative error, which turns into $O(m\eps)$ additive error since we only care about approximating up to $2m$. Hence the overall additive error is $O(m\eps)$, which can be made $m\eps$ by lowering the value of $\eps$.
    \end{proof}

Now we show that \cref{lem:greedy} can be used to solve \cref{probknap}, which is sufficient for proving \cref{theo:main}.
\begin{proof}[Proof of \cref{theo:main}]
    To solve \cref{probknap}, we divide $[1, 2\eps ^ {-1})$ into $O(\log (1/\eps))$ many intervals $[m,2m)$ where $m$ are powers of 2, and use \cref{lem:greedy} to obtain profit functions achieving $m\eps$-additive approximation up to $2m$. Then, taking their pointwise minima yields an $(1-O(\eps))$ approximation.
\end{proof}

In the following sections, we will prove \cref{lemma:knapmain}.

\subsection{Approximation using \texorpdfstring{$\Delta$}{Delta}-multiples of small set \texorpdfstring{$\Delta$}{Delta}}

We first introduce several additional tools borrowed from previous works that will be used in our final proof of \cref{lemma:knapmain}.

Following \cite{chan2018approximation}'s terminology, we say a monotone step function is \textit{$p$-uniform} if its function values are $0, p, 2p, \dots , lp$ for some $l$.
A $p$-uniform function is said to be \textit{pseudo-concave}, if the differences of consecutive $x$-breakpoints are nondecreasing from left to right. 
An example of a $p$-uniform and pseudo-concave function is the profit function $f_I$ of a set $I$ of items with the same profit $p_i=p$, which
 can be exactly computed by simple greedy: the function $f_I$ takes values $0,p,2p,\dots,np$, with $x$-breakpoints $w_1,w_1+w_2,\,\dots,\,w_1+\dots+w_n$, where $w_i$'s are sorted in nondecreasing order.

As in \cite{chan2018approximation} and \cite{DBLP:conf/icalp/Jin19}, we will use the method of approximation  via $\Delta$-multiples.
For a set $\Delta$ of numbers, we say that $p$ is a $\Delta$-multiple if it is a multiple of $\delta$ for some $\delta \in  \Delta$. 
Chan \cite{chan2018approximation} used the SMAWK algorithm \cite{aggarwal1987geometric} and suitable rounding to prove the following lemma:
\begin{lem}[\text{\cite[Lemma 5]{chan2018approximation}}] 
\label{improvedsmawk}
Let $f_1,\dots , f_m$ be monotone step functions with ranges contained in $[0, B]$. 
Let $\Delta  \subset [\delta, 8\delta]$. If every $f_i$ is $p_i$-uniform and pseudo-concave for some
$p_i \in [1, 2]$ which is a $\Delta$-multiple, then we can compute a monotone step function that $O(|\Delta|\delta)$-additively approximates 
$\min \{f_1 \oplus \dots \oplus f_m, B\}$ in $\tilde O( Bm/\delta)$ time. 
\end{lem}

        Chan \cite{chan2018approximation} gave a construction of a small set $\Delta$ such that every real number in $[1,2]$ can be approximated by a $\Delta$-multiple. Here, we present a more simplified construction.
        \begin{lem} \label{construction_delta}
        For parameters $0 < \epsilon < \delta < 1/2$, let $r=\lceil \log_{1+\epsilon} (1+2\delta) \rceil=O(\delta/\eps)$, and define $a_i=\delta(1+\eps)^i$ for $0\le i\le r+1$. Let $\Delta=\{a_i\}$ be the set of $a_i$. Then for any $t\in [1, 2]$, there is a multiple of some $a_i$ in the range $[t, t+2\epsilon]$. Thus, every real number in $[1, 2]$ can be approximated by a $\Delta$ multiple with $O(\eps)$ additive error, where $|\Delta|=r+2=O(\delta/\eps)$ and all elements in $\Delta$ are within $[\delta, 8\delta]$.  
        \end{lem}
        
        \begin{proof}
        Let $c$ be the largest integer such that $(t+2\eps)/c \ge a_0$. Since $c$ is largest, $c+1 \ge (t+2\eps)/a_0\ge 1/\delta$, so $(c+1)/c \le (1/\delta)/(1/\delta-1)=1+\delta/(1-\delta) \le 1+2\delta \le a_{r+1}/a_0$. Since $(t+2\eps)/(c+1) < a_0$, we know $(t+2\eps)/c < a_{r+1}$. Let $k$ be the largest integer in $[0, r]$ such that $a_k \le (t+2\eps)/c$. Note $a_{k+1} > (t + 2\eps)/c$, so $a_k=a_{k+1} / (1+\eps) \ge t/c$ using the fact that $t \le 2$. As a result, $a_k \in [t/c, (t+2\eps)/c]$, thus $ca_k \in [t, t+2\eps]$. 
        \end{proof}

    \subsection{Random Partitioning}
        Assume that $n<1/\epsilon$. In the section, we will use random partitioning to prove \cref{lemma:knapmain}, restated below. 
        
        \knapsackmainrestate*
        \begin{proof}

        Set $\Delta_1 = \Theta(\sqrt n) $ and $\Delta_0 = \Theta(n ^ \frac{7}{10}\epsilon ^ {\frac{2}{5}}2^{c\sqrt{\log (1/\eps)}})$ for some small constant $c>0$. Assume that $\Delta_0$ is a power of $2$ without loss of generality. Note that $\Delta_0 = O(\Delta_1)$, which follows from $n = O(1 / \eps)$.
        
        \begin{claim} \label{partG}
        We can partition elements of $I$ into $\Theta(\Delta_1)$ groups $G_1, G_2, \dots, G_k$, each of size $O(n/\Delta_1)$, while all elements within group $G_i$ are $(1+\eps)$-approximated by multiples of $p_i$ for some $p_i=\Theta(\Delta_1 \epsilon)$. 
        \end{claim}
        
        \begin{proof}
        
        In \cref{construction_delta}, plugging in $\delta=\eps \Delta_1$, we obtain a set $A$ of size $O(\Delta_1)$ whose elements are of order $\Theta(\Delta_1 \epsilon)$, and each item in $I$ can be $(1 + \epsilon)$-approximated by $A$-multiples. 
        
        We group the elements in $I$ by their divisor in $A$. We then evenly split groups with size more than $n/\Delta_1$ into two until all groups have sizes of at most $n/\Delta_1$.
        \end{proof}
        
        From now on, assume that $G_1, G_2, \dots, G_k$ are groups satisfying conditions in \cref{partG}.
        
        We now randomly partition $\{1, 2, \dots, k\}$ into $\Delta_0$ parts, $I_1, \dots, I_{\Delta_0}$, by assigning each $1\le i\le k$ into some $I_j (1\le j\le \Delta_0)$ independently and uniformly.
        Then, set $X_j=\bigcup_{i\in I_j} G_i$ for every $1\le j\le \Delta_0$. It is easy to see $\{X_j\}$ is a partition of $I$. 
        
        \begin{claim} \label{partX}
        With probability $\ge 3/4$, $|I_j|=O(\Delta_1/\Delta_0)$, and hence $|X_j| \le O(n/\Delta_0)$. 
        \end{claim}
        
        \begin{proof}
        By Chernoff bound\footnote{For independent random variables $x_1, \dots, x_n\in \{0, 1\}$ and $\delta>0, 0 \le w_1, \dots, w_n \le 1$, let $X=\sum_{i=1}^n w_ix_i$ and $\mu=\mathbb E[x]$, then $\Pr[|x-\mu|\ge \delta\mu] \le 2e^{-\delta^2\mu/3}$.}, for some large constant $c>0$, $|I_j| \ge ck/\Delta_0$ happens with probability at most $1/(4\Delta_0)$. Thus $|I_j|=O(\Delta_1/\Delta_0)$ holds for all $j$ with probability $\ge 3/4$ by union bound. By \cref{partG}, $|X_j| \le |I_j|O(n/\Delta_1)=O(m/\Delta_0)$. 
        \end{proof}
        
        Now assume the event in \cref{partX} happens. 
        
        \begin{claim} \label{innerTime}
        We can approximate $\bigoplus_{x\in X_j} f_x$ with additive error $O(n\eps/\Delta_0)$ for all $1\le j\le \Delta_0$ in \\ $\tilde O(n^2\eps^{-1}/(\Delta_0\Delta_1))=\tilde O(n ^ {{4}/{5}}\epsilon ^ {-{7}/{5}}/2^{\Omega(\sqrt{\log (1/\eps)})})$ time.
        \end{claim}
        
        \begin{proof}
        Fix a single $j$. By \cref{partX}, $\bigoplus_{x\in X_j} f_x$ is the convolution of $O(n/\Delta_0)$ elements, each being a multiple of order $\Theta(\Delta_1 \epsilon)$. By applying \cref{improvedsmawk} with parameters $B=O(n/\Delta_0), \delta=\Theta(\Delta_1 \epsilon), |\Delta|=|I_j|=O(\Delta_1/\Delta_0)$, we can approximate $\bigoplus_{x\in X_j} f_x$ with additive error $O(\Delta_1^2\eps/\Delta_0)=O(n\eps/\Delta_0)$ within time $\tilde O((n/\Delta_0)^2/(\Delta_1 \epsilon))$. 
        
        We can do so for all $1\le j\le \Delta_0$, with running time $\tilde O(n^2/(\Delta_0\Delta_1 \epsilon))=\tilde O(n ^ {{4}/{5}}\epsilon ^ {-{7}/{5}}/2^{\Omega(\sqrt{\log (1/\eps)})})$.
        \end{proof}
        
        Now we can replace $\bigoplus_{x\in X_j} f_x$ by the approximation obtained in \cref{innerTime}, since the total additive error inflicted will be $O(n\eps/\Delta_0)\Delta_0=O(n\eps)$.
        
        We use divide and conquer to combine the answer of $\bigoplus_{x\in X_j} f_x$. The merge process can be viewed as a complete binary tree with $\Delta_0$ leaves. For $S\subseteq \{1, 2, \dots, \Delta_0\}$, define $F(S)=\bigoplus_{x\in \cup_{(s\in S)} X_s} f_x$. \cref{innerTime} allows us to approximate $F(S)$ for all $|S|=1$. Now we have the following claim regarding combining two subtrees $S_1$ and $S_2$. 
        
        \begin{claim} \label{divconc}
        Assume $i\le \log_2 \Delta_0$ and $|S_1|=|S_2|=2^i$, where $S_1, S_2\subseteq \{1,2,\dots,\Delta_0\}$ and $S_1\cap S_2=\emptyset$. Assume that $A_1$ is an approximation of $F(S_1)$ with additive error $err_1$, $A_2$ is an approximation of $F(S_2)$ with additive error $err_2$.
        Then with probability $\ge 1-1/(5\Delta_0)$, we can compute an approximation of $F(S_1\cup S_2)$ with additive error $err_1+err_2+O(2^{0.9i}n\eps/\Delta_0)$ in time $O(\eps^{-2}\Delta_0^{0.5}/(\Delta_1^{0.5} 2^{\Theta(\sqrt{\log (1/\eps)})}))$.  
        \end{claim}
        
        \begin{proof}
        Define $\delta_i=2^{0.9i}n\eps/\Delta_0$. A naive way to approximate $F(S_1\cup S_2)$ is to round each value in $A_1$ and $A_2$ to a multiple of $\delta_i$, and then invoke the $(\min, +)$ convolution as in \cref{dc}. 
        In the following we will show a better method exploiting the fact that $\{X_i\}$ is a random partition. 
        
        Let the global optimal solution be to choose the subset $T$ of items. Define $H_1=\bigcup_{i\in S_1} X_i, H_2=\bigcup_{i\in S_2} X_i$. Note that the groups $G_1, \dots, G_k$ are assigned into $X_1, \dots, X_{\Delta_0}$ uniformly randomly. Pick $u=Cn\sqrt{2^i/(\Delta_1\Delta_0)}\log n$ for a large constant $C>0$. By Chernoff bound, the probability that $\Pr(|\sum_{x\in T\cap H_1} x - \sum_{x\in T\cap H_2} x| \ge u) \le 1/(5n)$.\footnote{We apply Chernoff bound with $w_j=R_j/(2n/\Delta_1)$ where $R_j =\sum_{x\in T\cap G_j} x$. Now consider $S_1$. We set $x_j=1$ if $j \in \cup_{t\in S_1} I_t$ and $x_j=0$ otherwise. Since the partition $\{I_t\}_{1\le t\le \Delta_0}$ is random, the expected value of $\sum_{j=1}^kw_jx_j$ will be $\Theta(\Delta_12^i/\Delta_0)$. From Chernoff bound, this value will be $u/(4n/\Delta_1)$ away from expected value with probability $\le 2e^{\Theta(-(u/(4n/\Delta_1))^2/(\Delta_12^i/\Delta_0))} \le 1/(10n)$. By union bound, both $\sum_{x\in T\cap H_2} x$ and $\sum_{x\in T\cap H_1} x$ will be within difference $u/2$ from the expected value with probability $\ge 1-1/(5n)$, in which case their difference will be bounded by $u$. } 
        
        Now assume that $|\sum_{x\in T \cap H_1} x - \sum_{x\in T \cap H_2} x| \le u$, and we show how to approximate $F(S_1\cup S_2)$ under the assumption. During the $(\min, +)$ convolution, we first round the values of $F(S_1),F(S_2)$ to multiples of $\delta_i$. Then we only need to consider the pairs that differ in value by at most $u$. We then divide the arrays into blocks with values within a difference of $u$ from each other, and do $(\min, +)$-convolution between the pairs of blocks with indices differing by at most 1.
        The block sizes are at most $u / \delta_i = \tilde O(\eps ^ {-1} \Delta_0 ^ {0.5} \Delta_1 ^ {-0.5} / 2 ^ {0.4i})$, so the running time for each $(\min, +)$-convolution is $O(\eps ^ {-2} \Delta_0 / (\Delta_1 2 ^ {0.8i} 2 ^ {\Omega(\sqrt{\log(1 / \epsilon)})}))$ using Williams's $O(n ^ 2 / 2 ^ {\Omega(\sqrt{\log(1 / \epsilon)})})$-time algorithm for length-$n$ $(\min, +)$-convolution \cite{williams2014faster}. Since the value in the merged answer is bounded by $\tilde O(2^in/\Delta_0)$ by \cref{partX}, there are $\tilde O(2^in/(\Delta_0u))$ min-plus convolutions in total, with total complexity $\tilde O(\eps^{-2}2^in/ (\Delta_1 2 ^ {0.8i} 2 ^ {\Omega(\sqrt{\log(1 / \epsilon)})}u))=O(\eps^{-2}\Delta_0^{0.5}/(\Delta_1^{0.5} 2^{\Theta(\sqrt{\log (1/\eps)})}))$.
        \end{proof}
        
        Now we conclude the proof by applying \cref{divconc} to the divide and conquer process. Assume all the $\le \Delta_0$ many calls to \cref{divconc} yield correct approximations, which happens with success probability $\ge 3 / 4$ by union bound.
        
        To analyze the error term, note that there are $O(\Delta_0/2^i)$ merges of two subtrees with $2^i$ parts each, where \cref{divconc} inflicts additive error $O(2^{0.9i}n\eps/\Delta_0)$ for each such a merge. Thus the total additive error is bounded by $\sum_{2^i\le \Delta_0} (2^{0.9i}n\eps/\Delta_0)(\Delta_0/2^i)=O(n\eps)$.
        
        Now we analyze the time complexity. Note that the total complexity for the $i$-th layer is
        \begin{align*}
                &O((\Delta_0 / 2^i) \cdot \eps ^ {-2} \Delta_0 ^ {0.5} \Delta_1 ^ {-0.5} / 2 ^ {\Omega(\sqrt{\log(1 /  \epsilon)})}) \\
             = \  &O(\eps ^ {-2} \Delta_0 ^ {1.5} \Delta_1 ^ {-0.5} / 2 ^ {\Omega(\sqrt{\log(1 /  \epsilon)})}) \\
             = \ & O(n ^ {\frac{4}{5}}\epsilon ^ {-\frac{7}{5}} / 2 ^ {\Omega(\sqrt{\log(1 /  \epsilon)})}).
        \end{align*}
        As there are logarithmically many layers, the total complexity for the divide and conquer part is $O(n ^ {\frac{4}{5}}\epsilon ^ {-\frac{7}{5}} / 2 ^ {\Omega(\sqrt{\log(1 /  \epsilon)})})$.
        
        Thus our total complexity is  $O(n ^ {\frac{4}{5}}\epsilon ^ {-\frac{7}{5}} / 2 ^ {\Omega(\sqrt{\log(1 /  \epsilon)})})$, and with a success probability of $\ge 1/2$ (which can be amplified by repetition) by union bound. 
        
        A small detail is that when $n$ is so small that $n ^ {7 / 10} < \eps ^ {-2 / 5}$, $\Delta_0 < 1$ and our reasoning falls apart. In such cases, one can simply set $\Delta_0 = 1$ and the running time still holds.
        \end{proof}

        \section{Approximating Partition}

\label{sec:partition}

In this section, we will solve the following problem.
\begin{restatable}{prob}{prorestate}
\label{prob1}
Assume $\eps\in (0,1/2)$ and $1/\eps \in \N^+$. Given a set $X$ of $n$ \emph{distinct} integers in the interval $[1/\eps,2/\eps)$, compute a set $A\subset\N$ that $n$-additively approximates $\caS(X)$. 
\end{restatable}
By a tedious reduction that is heavily based on known techniques, one can show the following.
\begin{restatable}{lem}{restateredpart}
	\label{lem:reduction} If for some $c\ge 1$, \cref{prob1} can be solved in $\tilde O(n + 1/\eps^c)$ time, then $(1-\eps)$-approximating \emph{\parti} can also  be solved in $\tilde O(n + 1/\eps^c)$ time. 
\end{restatable}
\cref{lem:reduction} will be proved in the appendix.

Now we proceed to  describe our main algorithm for solving \cref{prob1}.

In the following lemma, we merge the approximations of $\caS(X_1),\caS(X_2)$ and obtain an approximation of $\caS(X_1\uplus X_2)$. When $X_1,X_2$ come from a short interval $[\ell,\ell+d]$, we can use densification via 2D FFT to obtain a speedup over the straightforward algorithm. 

\begin{lem}\label{lem:2dfft-add}
Let $\delta \in (0,1/2)$, and $\ell,d,t,\Delta \in \N^+$ such that $d\le \ell \le t$.

Let $X_1,X_2 \subseteq \N^+ \cap [\ell, \ell+d]$ be two integer sets. 
Given $A_1,A_2 \subset \N$ as input where for $i\in \{1,2\}$, $A_i$ is an  $(1-\delta)$ approximation of $\caS(X_i)$ up to $t$, one can compute a set $A \subset \N^+$ of size $|A|\le Z$
that $(1-\delta,\Delta-1)$-approximates $\caS(X_1\uplus X_2)$ up to $t$,
in $\tilde O(Z+|A_1|+|A_2|)$ time, where
\[ Z \le  O \left ( \min \left \{ \left \lceil \frac{t}{\Delta} \right \rceil,\,\, \frac{t}{\ell} \cdot \left \lceil \frac{t d}{\ell \Delta} \right \rceil  \right \}\right ).\]
\end{lem}

\begin{proof}
Let $\bar \Delta := \lceil \Delta/2\rceil$.
We will run one of the following two algorithms that minimizes $Z$.

\paragraph*{Algorithm 1 (1D FFT).}
For $i\in \{1,2\}$, by rounding the integers in $A_i$ down to multiples of $\bar \Delta$, we obtain set $A_i' \subset \bar \Delta\cdot  \N $  that $(\bar \Delta-1)$-additively approximates $A_i$. Then, since $A_i' \subseteq [0,t]$, their sumset $A_1'+A_2'$ can be computed by FFT in $\tilde O(\lceil t/\bar \Delta \rceil ) \le \tilde O(\lceil t/ \Delta \rceil )$ time.
Note that $A:=A_1'+A_2'$ approximates $A_1+A_2$ with additive error at most $2 (\bar \Delta -1) \le \Delta -1$, so $A$ is a $(1-\delta, \Delta-1)$-approximation of $\caS(X_1 \uplus X_2)$ up to $t$, due to \cref{prop:merge}.
\paragraph*{Algorithm 2 (Densification with 2D FFT).}
For every $a\in A_i$, there exists $s \in \caS(X_i;t)$ such that $0 \le s-a \le s\delta$. Note that $s$ is the sum of at most $ t/\ell $ many integers from $[\ell,\ell+d]$, so $s$ can be expressed as $s = k\ell + b'$ for some $k\in \N \cap [0,t/\ell]$ and $0 \le b'\le  dt/\ell$. Hence, $a\in A_i$ can be expressed as $a = k\ell + b$ for some $k\in \N \cap [0,t/\ell]$ and $  -s\delta\le b\le  dt/\ell $.
Then, by rounding $b$ down to integer multiples of $\bar \Delta $, we obtain $A'_i \subset \N$  that ($\bar \Delta -1$)-additively approximates $A_i$, such that every $a' \in A'_i$ can be expressed as
\[a' = k\ell + j \bar \Delta,\]
for some $k \in \N \cap [0,t/\ell]$ and $j \in \Z \cap [ -1-s\delta/\bar \Delta , dt/(\ell \bar \Delta )]$.
Using this 2-dimensional $(k,j)$ representation of $A'_i$, we can  compute $A'_1+A'_2$ using 2D FFT (\cref{lem:2dfft}): the first dimension has size $O(t/\ell)$, and the second dimension has size at most 
\begin{align*}
   dt/(\ell\bar \Delta ) + s\delta/\bar \Delta + O(1) \le O \left (\left \lceil \frac{t d}{\ell \Delta} \right \rceil \right ),
\end{align*}
where the inequality follows from  $s\le t$ and an assumption 
\begin{equation}
\label{eq:assump}
    \delta \le O( d/\ell),
\end{equation} which will be justified later. Hence, the running time of this 2D FFT is \[\tilde O\left ( \frac{t}{\ell} \cdot \left \lceil \frac{t d}{\ell \Delta} \right \rceil \right ).\]

Similarly to Algorithm~1, one also can show that in this case $A:=A_1'+A_2'$ is a $(1-\delta, \Delta-1)$-approximation of $\caS(X_1 \uplus X_2)$ up to $t$.

To justify assumption \eqref{eq:assump}, observe that if $\delta \ge d/(\ell+d)$ holds instead, or equivalently, $(1-\delta) (\ell+d)\le \ell$, then one can round every integer in $X_1,X_2 \subset [\ell,\ell+d]$ down to exactly $\ell$ while still ensuring $(1-\delta)$ approximation, and hence immediately obtain an $A \subset \ell \cdot  \N$ of size $|A| \le \lceil t/\ell \rceil$ that $(1-\delta)$-approximates $\caS(X_1\uplus X_2)$ up to $t$.
\end{proof}

We then apply 
\cref{lem:2dfft-add} with scaling, and obtain the following lemma that has purely multiplicative approximation.
\begin{lem}
\label{lem:2dfft-mult}
Let $\delta,\delta_0\in (0,1/2)$, and $\ell,d,T \in \N^+$ such that $d\le \ell \le T$.

Let $X_1,X_2 \subseteq \N^+ \cap [\ell, \ell+d]$ be two integer sets. 
Given $A_1,A_2 \subset \N$ as input where for $i\in \{1,2\}$, $A_i$ is an  $(1-\delta)$ approximation of $\caS(X_i)$ up to $T$,
 one can compute a set $A \subset \N^+$ of size $|A|\le Z$
that $(1-\delta-\delta_0)$-approximates $\caS(X_1\uplus X_2)$ up to $T$,
in $\tilde O\big (Z+(|A_1|+|A_2|)\log (2T/\ell)\big )$ time, where
\[ Z \le 
  O\left (\min \left \{ \frac{\log (2T/\ell)}{\delta_0},\,\,  \frac{T}{\ell} \cdot \left \lceil \frac{ d}{\ell \delta_0} \right \rceil   \right \}\right ) . \]
\end{lem}
\begin{proof}
Initialize  set $A = \{0\}$.
We iterate over all $r$ being integer powers of $2$ such that $\ell/6 \le r\le T$.
For each $r$, apply \cref{lem:2dfft-add} to $A_1$ and $A_2$ with $t:= 6r$ and $\Delta:= \lceil \delta_0 r\rceil $, and obtain a set $A_r\subseteq \N\cap [0,6r]$ that $(1-\delta,\lceil \delta_0 r\rceil-1)$-approximates  $\caS(X_1 \uplus X_2)$ up to $6r$. We then insert all elements in $A_r \cap [r,6r]$ into $A$.
We will show that eventually $A$ is a $(1-\delta_0-\delta)$-approximation of $\caS(X_1 \uplus X_2)$ up to $T$.

Observe that for every $a\in A_r \cap [r,6r]$, there exists $s\in \caS(X_1\uplus X_2)$ such that $a\le s$ and 
\begin{align*}
    a&\ge (1-\delta) s - ( \lceil \delta_0 r\rceil -1) \\ 
    &> (1-\delta) s - \delta_0 r\\
    & \ge (1-\delta-\delta_0 ) s, 
\end{align*}
where the last step follows from $s\ge a \ge r$.

Conversely, for every positive $s \in \caS(X_1\uplus X_2; T) $ (which must satisfy $\ell \le s\le T$), let $r$ be a power of two such that $3r\le s \le 6r$. Then there exists $a\in A_r$ such that $a\le s\le 6r$ and 
\begin{align*}
    a &\ge (1-\delta) s - ( \lceil \delta_0 r\rceil -1)\\
     & \ge (1-\delta ) s - \delta_0 r\\
     & \ge s/2 - r/2\\
     & \ge r,
\end{align*}
so $a \in A_r \cap [r,6r]$ and hence will be included in $A$, and similarly as before we have $a\ge (1-\delta_0-\delta)s$. Hence, we have established that $A$ is a $(1-\delta_0-\delta)$-approximation of $\caS(X_1 \uplus X_2)$ up to $T$.

It remains to bound the total running time and the size of $A$. There are $O(\log (2T/L))$ many iterations of $r$, where for each $r \in [\ell/6, T]$ with $t:=6r$ and $\Delta:= \lceil \delta_0 r\rceil$, \cref{lem:2dfft-add} gives the upper bound
\begin{align*}
    Z_r &\le  O \left ( \min \left \{ \left \lceil \frac{t}{\Delta} \right \rceil,\,\, \frac{t}{\ell} \cdot \left \lceil \frac{t d}{\ell \Delta} \right \rceil  \right \}\right )\\
    &  \le   O \left ( \min \left \{ \left \lceil \frac{r}{\delta_0 r} \right \rceil,\,\, \frac{r}{\ell} \cdot \left \lceil \frac{r d}{\ell \delta_0 r} \right \rceil  \right \}\right )\\
    & \le O \left ( \min \left \{ \left \lceil \frac{1}{\delta_0} \right \rceil,\,\, \frac{r}{\ell} \cdot \left \lceil \frac{ d}{\ell \delta_0} \right \rceil  \right \}\right ).
\end{align*} 
Hence, summing over all powers of two in the range $[\ell/6,T]$, we have
\[ Z \le \sum_r Z_r \le O\left (\min \left \{ \frac{\log (2T/\ell)}{\delta_0}, \,\, \frac{T}{\ell} \cdot \left \lceil \frac{ d}{\ell \delta_0} \right \rceil   \right \}\right ) . \qedhere
 \]
\end{proof}

\Cref{lem:2dfft-mult} implies the following immediate corollary by dropping the upper bound $T$.
\begin{cor}
\label{cor:merge}
Let $\delta,\delta_0\in (0,1/2)$, and $\ell,d \in \N^+$ such that $d\le \ell $.

Let $X_1,X_2 \subseteq \N^+ \cap [\ell, \ell+d]$ be two integer sets of total size $|X_1|+|X_2|=n$. 
Given $A_1,A_2 \subset \N$ as input where for $i\in \{1,2\}$, $A_i$ is an  $(1-\delta)$ approximation of $\caS(X_i)$,
 one can compute a set $A \subset \N^+$ of size $|A|\le Z$
that $(1-\delta_0-\delta)$-approximates $\caS(X_1\uplus X_2)$,
in $\tilde O\big (Z+(|A_1|+|A_2|)\log n\big )$ time, where
\[ Z \le  O\left (\min \left \{ \frac{1}{\delta_0},\,\,   \frac{ nd}{\ell \delta_0}+ n   \right \}\cdot \log n\right ) . \]
\end{cor}
\begin{proof}
Immediately follows from \cref{lem:2dfft-mult} by setting $T = n\cdot (\ell+d) $, which is an upper bound on the largest element of $\caS(X_1\uplus X_2)$.
\end{proof}

Now, we apply \cref{cor:merge} in a divide-and-conquer fashion, to approximate the subset sums of $X\subseteq \N^+ \cap [\ell,2\ell]$.
\begin{lem}\label{lemma:dc}
Let $\delta \in (0,1/2)$ and $\ell \in \N^+$.

Given an integer set $X\subseteq \N^+ \cap [\ell,2\ell]$ of $n$ integers, one can compute a set $A\subset \N^+$ that $(1-\delta)$-approximates $\caS(X)$, in $\tilde O(n + \sqrt{n}/\delta)$ time.
\end{lem}
\begin{proof}
Let $X = \{x_1,x_2,\dots,x_n\}$ where $\ell \le x_1<x_2<\dots<x_n\le 2\ell$. Set $\delta_0: = \delta/ \lceil \log_2 n\rceil$.

We  will use a divide-and-conquer approach to merge the items of $X$ using \cref{cor:merge}. Build a balanced binary tree with $n$ leaf nodes representing the items $x_1,\dots,x_n$ from left to right. At each internal node representing $x_{[l..r]}$, we use \cref{cor:merge} to merge the results of the two child nodes (representing $x_{[l..m]}$ and $x_{[m+1..r]}$ respectively, where $m=\lfloor (l+r)/2\rfloor$), and obtain an approximation of $\caS(\{x_l,x_{l+1},\dots,x_{r}\})$. Finally we obtain an approximation of $\caS(X)$ at the root node.

The binary tree has $\lceil \log_2 n\rceil $ levels, where each level of applying \cref{cor:merge} worsens the approximation factor by $\delta_0$. Hence, the overall approximation factor of $\caS(X)$ is $1-\delta_0\cdot \lceil \log_2 n\rceil = 1-\delta$ as required.

It remains to bound the total running time of all invocations of \cref{cor:merge}. Note that in each invocation, the $(|A_1|+|A_2|)\log n$ summand in the stated time complexity is always absorbed (up to $\log n $ factors) by the output sizes of the two child nodes, which are in turn bounded by the running times of these two child invocations. So it suffices to bound the sum of the $Z$ quantity stated in \cref{cor:merge}.

We separately bound for each level of the binary tree. At the $i$-th level $(0\le i < \lceil \log_2 n\rceil)$, there are at most $m_i= 2^i$ invocations of \cref{cor:merge}, where each invocation involves at most $n_i = \lceil n/2^i\rceil $ items in $X$.  Note that $n_im_i \le 2n$. Suppose these $m_i$ invocations involve $x_{[1..k_1]},x_{[k_1+1..k_2]},\dots,$ $x_{[k_{m_i-1}+1..n]}$ respectively. Then the $j$-th invocation has $d$ value (stated in \cref{cor:merge}) at most $d_j \le x_{k_j}-x_{k_{j-1}}$. Hence, the sum of these $d$ values is at most \begin{equation}
    \sum_{j=1}^{m_i}d_j \le \sum_{j=1}^{m_i}(x_{k_j} - x_{k_{j-1}}) \le x_{n} - x_1 \le \ell.
    \label{eq:sumbound}
\end{equation}

Now we are ready to bound the sum of the $Z$ quantity over the $m_i$ invocations at level $i$  $(0\le i < \lceil \log_2 n\rceil)$. We consider two cases.
\begin{itemize}
    \item \textbf{Case 1}: $n_i\le \sqrt{n}$.
    
    Then, by \cref{cor:merge},
    \begin{align*}
        \sum_{j=1}^{m_i} Z_j &\le \sum_{j=1}^{m_i} \left (\frac{n_i d_j}{\ell\delta_0}+n_i \right ) \cdot \log n\\
        & =  \left (\frac{n_i\sum_{j=1}^{m_i}d_j}{\ell \delta_0}+m_in_i \right )  \cdot \log n\\
        & \le \left (\frac{n_i}{\delta_0}+m_in_i \right )  \cdot \log n \tag{by \eqref{eq:sumbound}}\\
        & \le \tilde O\left ( \frac{\sqrt{n}}{\delta_0}  + n\right ).
    \end{align*}
    \item \textbf{Case 2:} $n_i> \sqrt{n}$. 
    
    Then, $m_i \le 2n/n_i < 2\sqrt{n}$. By \cref{cor:merge},
    \begin{align*}
        \sum_{j=1}^{m_i} Z_j &\le m_i \cdot \frac{1}{\delta_0}\cdot \log n\\
        & \le \tilde O(\sqrt{n}/\delta_0).
    \end{align*}
\end{itemize}
Hence, in either case we have $\sum_{j=1}^{m_i} Z_j \le \tilde O(n+ \sqrt{n}/\delta)$. Hence, the total running time over all levels $0\le i < \lceil \log_2 n\rceil$ is also $\tilde O(n+ \sqrt{n}/\delta)$.
\end{proof}

Finally, we solve \cref{prob1} by combining \cref{lemma:dc} with the additive combinatorics results of  \cite{DBLP:journals/siamcomp/GalilM91,bw21}.

\begin{lem} \label{lemma:main}
We can solve \cref{prob1} in $\tilde O\left ( n +  \min \{\eps^{-1} n^{1/2},\,\, \eps^{-1} +\eps^{-2}/n^{3/2}\}  \right )$ time, which is at most $\tilde O(n+ 1/\eps^{5/4})$ .
\end{lem}

\begin{proof}
Recall that in \cref{prob1}, for $\eps>0$ where $\ell =  1/\eps$ is an integer, we are given a set $X\subseteq \N^+ \cap [\ell,2\ell)$ of $n$ distinct integers,  and need to compute a set $A\subset\N$ that $n$-additively approximates $\caS(X)$. 

We will choose to run one of the following two algorithms depending on the parameters.

\paragraph*{Algorithm 1.}
Directly apply \cref{lemma:dc} with $\delta:=\eps$, in $\tilde O(n+\sqrt{n}/\eps)$ time.

When $n\le \tilde O(1/\eps^{1/2})$, the running time of Algorithm~1 is $\tilde O(n+1/\eps^{5/4})$.
\paragraph*{Algorithm 2.}
Let $\sigma = \Sigma(X)$, and let $\lambda$ be the threshold value from Theorem \ref{theo:galil} satisfying  $\lambda = \tilde \Theta(\ell^2/n)$. The following algorithm applies when $\lambda \le \sigma/2 $, which holds in particular when $1/\eps \ll n^2$. 

Initialize $A=\emptyset$. We set $\delta:= n/(n+\lambda)$, and apply \cref{lemma:dc} in $\tilde O(n+ \sqrt{n}/\delta)$ time to compute a set $A_\delta$ that $(1-\delta)$-approximates $\caS(X)$. 
Observe that $A_\delta \cap [0, \lambda]$ is an $n$-additive approximation of $\caS(X)$ up to $\lambda$. Hence, we insert all elements in $A_\delta \cap [0, \lambda]$ to $A$. 

Then, using the data structure from \cref{theo:galil}, we compute an $n$-additive approximation of $\caS(X) \cap [\lambda,\sigma/2]$ and insert them into $A$. To do this, we start from the left endpoint $\lambda$ of the interval $[\lambda,\sigma/2]$, and each time use binary search (implementable using range queries supported by \cref{theo:galil}) to find the next subset sum in the interval, and then jump $n$ steps to the right since we allow an additive error of $n$. The time complexity is $O(\lceil \frac{\sigma/2-\lambda}{n}\rceil \cdot \log \sigma) \le \tilde O(\ell)$.

Now we have constructed $A$ that $n$-additively approximates $\caS(X)$ up to $\sigma/2$. Using the simple fact that $t \in \Sigma(X)$ if and only if $\sigma-t\in \Sigma(X)$, we can symmetrically use $A$ to obtain an approximation of the remaining half. Specifically, letting $A':= \{\sigma - a-n: a\in A\}$, it is straightforward to verify that $A\cup A'$ is an $n$-additive approximation of $\caS(X)$  (up to $\sigma$). So we return $A\cup A'$.

The overall time complexity of Algorithm~2 is 
\begin{align*}
    \tilde O(n + \ell + \sqrt{n}/\delta) &\le \tilde O\left (n+1/\eps + \frac{\sqrt{n}(n+\lambda)}{n}\right )\\
    & \le \tilde O\left (n+1/\eps + \frac{1/\eps^2}{n^{3/2}}\right ).
\end{align*} 
When $n \gg 1/\eps^{1/2}$, the running time is $\tilde O(n+1/\eps^{5/4})$.
\end{proof}

Combined with the reduction in \cref{lem:reduction}, this proves our main \cref{theo:parti}.

	\bibliographystyle{alphaurl} 
	\bibliography{main}

    \appendix
    
\section{Know reductions from the Knapsack Problem}

Recall that we defined the following simpler problem.
\proknaprestate*

\redknap*

\begin{proof}
First, we can reduce $\eps$ so that $1/\eps$ becomes an integer.

 We will restrict the profit values into small intervals, as follows:  divide the items into $O(\log \frac{\max_j p_j}{\min_j p_j}) =$ $O(\log \eps^{-1})$ groups (see \cref{sec:prelimknap}), each containing items with $p_i \in [2^{j},2^{j+1}]$ for some $j$ (which can be rescaled to $[1,2]$). Finally, use the merging lemma \cref{dc} to merge the profit functions of all groups, in $\tilde O(n+\eps^{-2})$ overall time.  
 
 Now, having restricted the profit values into $[1,2)$, we can  round every profit value to a multiple of $\eps$, which incurs only $(1-O(\eps))$ approximation factor in total.
 
 Finally, the following greedy lemma takes care of the case with total profit above $\Omega(\eps^{-1})$. 
\begin{lem}
\label{sortgreed}
 Suppose $p_i \in [1,2]$ for all $i\in I$. For $B=\Omega(\eps^{-1})$,  the profit function $f_I$ can be approximated with  additive error $O(\eps B)$ in $O(n\log n)$ time.
\end{lem}
\begin{proof}
Simply sort the items in nonincreasing order of efficiency $p_i/w_i$, and define the profit function $\tilde f$ resulting from greedy, with function values $0,p_1,p_1+p_2,\dots,p_1 + \dots + p_n$ and $x$-breakpoints $0,w_1,w_1+w_2,\dots,w_1+\dots+w_n$.  It clearly approximates $f_I$ with an additive error of $\max_i p_i \le 2 \le O(\eps B)$ for $B=\Omega(\eps^{-1})$. 
\end{proof}
This greedy approach achieves $(1-O(\eps))$-approximation for large profit values. Hence, it is sufficient to approximate $f_I$ up to $2/\eps$.
\end{proof}

\section{Proof of \cref{lemma:density}}
\label{app:density}

We need several results on dense subset sums developed by a series of works including \cite{sarkozy1994fine,lev2003blocks,DBLP:journals/siamcomp/GalilM91,bw21}.
The following definitions and theorems are from \cite{bw21}.   The sets considered here contain \emph{distinct positive integers}.

\begin{restatable}[Density]{definition}{defdensity} \label{df:den}
    A set $X \subset \N^+ $ is {\em $\delta$-dense} if it satisfies $|X|^2 \ge
    \delta\cdot \max {X}$.
\end{restatable}

\begin{restatable}[Almost Divisor]{definition}{defalmostdivisor} \label{df:ald}
    Let $X(d) := X \cap d \mathbb{Z}$ denote the set of all numbers in~$X$ that are
  divisible by $d$. 
  Let $\overline{X(d)} := X \setminus X(d)$ denote the
  set of all numbers in $X$ not divisible by $d$.
  We say an integer $d > 1$ is an {\em $\alpha$-almost divisor} of $X$ if $|\overline{X(d)}| \le
  \alpha \cdot  \Sigma(X) / |X|^2$.
\end{restatable}

        \begin{them}[{\cite[Theorem 4.1]{bw21}}]
            \label{thm:bw41}
  Let $\delta \ge 1$ and $ \alpha \le \delta/16$.
  Given an $\delta$-dense set $X$ of size $n$, there exists a  positive integer $d $ such that $X' := X(d)/d$ is $\delta$-dense and has no $\alpha$-almost divisor, and  the following additional properties are satisfied: 
  \begin{enumerate}
    \item $d \le 4  \Sigma(X) / |X|^2$, \label{item:my2}
    \item $|X'| \ge 0.75 |X|$, \label{item:my3}
    \item $\Sigma ({X'}) \ge 0.75\, \Sigma(X)/d$. \label{item:my}
  \end{enumerate}
        \end{them}
        \begin{them}[{\cite[Theorem 4.2]{bw21}}]
  Let $X$ be a multi-set and set
    \begin{align*}
    C_\delta &:= 1699200 \cdot \log (2n) \log^2(2), \\
    C_\alpha &:= 42480 \cdot \log(2), \\
    C_\lambda &:= 169920\cdot \log(2).
  \end{align*}
  If $X$ is $C_\delta$-dense and has no $C_\alpha$-almost divisor,
  then for $\lambda_{X} := C_\lambda \cdot  (\max {X}) \cdot \Sigma({X}) / |X|^2$ we have 
  \[\big  ( \, [\lambda_{X}, \Sigma({X}) - \lambda_{X}] \, \cap \Z \,\big ) \subseteq
  \caS(X). \]
  \label{thm:bw42}
        \end{them}
Now we are ready to prove \cref{lemma:density}.
        \densityrestate*

        \begin{proof}
            Let $C_{\delta} = \Theta(\log n), C_\alpha = \Theta(1), C_{\lambda}=\Theta(1)$ be defined as in \cref{thm:bw42}. Then, $X$ is $C_{\delta}$-dense since $C_{\delta}\cdot \ell = o(n^2)$. 
            Let $d$ be the positive integer guaranteed by \cref{thm:bw41} such that $X':=X(d)/d$ is $C_{\delta}$-dense and has no $C_{\alpha}$-almost divisor.

Then, by \cref{thm:bw42}, for 
$\lambda_{X'} := C_\lambda \cdot  (\max {X'}) \cdot \Sigma({X'}) / |X'|^2$ we have
  \[\big  ( \, [\lambda_{X'}, \Sigma({X'}) - \lambda_{X'}] \, \cap \Z \,\big ) \subseteq
  \caS(X'). \]
 Since $X'$ is $C_\delta$-dense, we have $\lambda_{X'}/\Sigma(X')  = C_\lambda \cdot  (\max {X'}) / |X'|^2 \le C_\lambda /C_{\delta}<0.1$.

  Now, let $\lambda:= d\cdot \lambda_{X'}$. From Property~\ref{item:my} in \cref{thm:bw41}, we know that 
  \[  \frac{\Sigma(X)/2}{d} \le  \frac{2}{3}\Sigma(X') < \Sigma(X')-\lambda_{X'}.\]  Hence, given any $\lambda \le t\le \Sigma(X)/2$,  we have \[\lambda_{X'} \le \lceil t/d\rceil \le \Sigma({X'}) - \lambda_{X'}.\]
  So $\lceil t/d\rceil \in \caS(X')$, which implies $t':=d\cdot \lceil t/d \rceil \in \caS(X)$. From Property~\ref{item:my2} in \cref{thm:bw41}, we have 
  \[ 0\le t'-t < d\le 4\cdot |X|\cdot (2\ell)/|X|^2 = 8\ell/n.\]
  Finally, we upper-bound $\lambda$ as
\begin{align*}
    \lambda &= d\cdot  C_\lambda \cdot  (\max {X'}) \cdot \Sigma({X'}) / |X'|^2 \\
&\le  d\cdot  C_\lambda \cdot  (\max {X'})^2  / |X'| \\
    & \le d\cdot C_\lambda \cdot (2\ell/d)^2/ |X'|\\
    & \le d\cdot C_\lambda \cdot (2\ell/d)^2/ (0.75n) \tag{by Property~\ref{item:my3}}\\
    & = (16C_\lambda/3)\cdot \frac{\ell^2}{dn}\\
    & \le O(\ell^2/n). \qedhere
\end{align*}
        \end{proof}

        \section{Known reductions from the Partition Problem}

Recall that we defined the following simpler problem.
\prorestate*
In the following, we will reduce Partition to this problem.

By a simple greedy argument, we can assume $\opt \ge t/2$.
\begin{lem}[e.g., {\cite[Lemma 4.3]{MuchaW019}}]
One may assume w.l.o.g.\ that for any \subs instance $\opt \ge t/2$. Otherwise the instance can be solved exactly in $\tilde O(n)$ time.
\label{lem:gred}
\end{lem}

Then, we have the following important lemma about $(1-\eps)$-approximating \parti. The key insight behind this lemma was first observed in \cite{MuchaW019}, indicating that approximating \parti is much easier than approximating general \subs instances.
\begin{lem}[c.f.\ \cite{MuchaW019}]
Let $X\subset \N^+$ be a multiset with sum of elements $\sigma=\Sigma(X)$, and let $\eps\in (0,1/2)$. Given a set $A\subset \N$ that $\eps\sigma /4$-additively approximates $\caS(X)$, one can immediately solve $(1-\eps)$-approximation \parti on $X$.
\label{lem:apx}
\end{lem}
\begin{proof}
Recall that $t=\sigma/2$, and $\opt = \max \{\Sigma(Y): \Sigma(Y)\le t, Y\subseteq X\}$. 

Given $A$, let $a:= \max\{a\in A: a\le t\}$. We claim that \[(1-\eps)\opt \le \min\{a, t(1-\eps/2) \} \le \opt,\]
which allows us to solve $(1-\eps)$-approximation \parti on $X$.

We prove this claim by separately considering two cases.
\begin{itemize}
    \item Case 1: $ a \le t(1-\eps/2)$.
    
     By definition of $A$, there exists $s\in \caS(X)$ such that $s - \eps\sigma/4\le a \le s$. We have $s\le a+\eps\sigma/4 \le t(1-\eps/2) + \eps \sigma/4 = t$, so $s\in \caS(X;t)$ and hence $\opt\ge s \ge a$. By \cref{lem:gred} we can assume  $t/2\le \opt \le t$. Then by definition of $A$ there exists $a' \in A$ such that $a'\le \opt \le t$ and $a'\ge \opt-\eps \sigma/4 \ge \opt - \eps \opt$. Then, by definition of $a$, we have $a\ge a' \ge (1-\eps)\opt$.
     
     Hence, we have established
     \[(1-\eps)\opt \le a=\min\{a, t(1-\eps/2) \} \le \opt.\]
     \item Case 2: $a > t(1-\eps/2)$.
     
       By definition of $A$, there exists $s\in \caS(X)$ such that $s - \eps\sigma/4\le a \le s$. We have $s\le a+\eps\sigma/4 \le t + \eps \sigma/4 = t(1+\eps /2)$, and $s\ge a > t(1-\eps/2)$. 
       
       By taking complement, we know $\sigma - s \in \caS(X)$ as well. Using the crucial fact that $t = \sigma/2$, we see that $\min \{s, \sigma-s\} \in \caS(X;t)$ and hence $\opt \ge \min \{s, \sigma-s\}$. Then, since $s\in (t(1-\eps/2),t(1+\eps/2)]$, we have $\min \{s, \sigma-s\}  \ge t(1-\eps/2)$.
       
     Hence, we have established
     \[(1-\eps)\opt \le t(1-\eps/2)=\min\{a, t(1-\eps/2) \} \le \opt. 
     \qedhere
     \]
\end{itemize}
\end{proof}
Using \cref{lem:apx}, we can solve $(1-\eps)$-approximation \parti by finding an additive approximation of $\caS(X)$. 

We are going to further simplify the input instance $X$. First we need the following lemma, which reduces the number of duplicate items in the input, by grouping them into powers of two.	 The proof of this lemma appeared in \cite{MuchaW019}, based on an earlier proof of a similar statement \cite[Lemma 2.4]{kx19}.
\begin{lem}[{\cite[Lemma 4.1]{MuchaW019}}]
\label{lem:reduce}
Given a multiset $S$ of $n$ integers from $[t]$, one can compute a multiset $T$ in $O(n \log n)$ time such that:
\begin{itemize}
    \item $\caS(S;t) = \caS(T;t)$.
    \item $|T|\le |S|$.
    \item  No element in $T$ has multiplicity exceeding two.
    \item For every $y \in T$, there is a corresponding $x\in S$ such that $y=2^k\cdot x$ for some $k\in \N$.
\end{itemize}
\end{lem}

Now we prove the main lemma.
\restateredpart*

	\begin{proof}
	 Let $X\subset \N^+$ be the input multiset of the $(1-\eps)$-\parti problem. We can without loss of generality assume $1/\eps$ is an integer.
	 
	Recall that $t = \sigma/2 = \Sigma(X)/2$. 
	We define a multiset $Y\subset \N^+$ as follows: for every $x\in X$, round $x$ down to the nearest integer multiple of $\lceil \frac{\sigma}{100n/\eps} \rceil$, denoted as $y$, and insert $y$ into $Y$ if $y$ is nonzero.
	Since the total incurred additive loss is at most $n\cdot (\lceil \frac{\sigma}{100n/\eps} \rceil-1) \le \frac{\eps\sigma}{100}$, we know that $\caS(Y)$ is an $\eps\sigma/100$-additive approximation of $\caS(X)$.
	
	Now,  we can without loss of generality assume $y\in [1/\eps,  100n/\eps^2]\cap \N^+$ for  all $y\in Y$, since otherwise we could simply scale all elements in $X,Y$  (as well as $\sigma,t$).
	
	Then, define another multiset $Z \subset \N^+$ as follows: for every $y\in Y$,  round $y$ down to $2^k \cdot z_0$ for some $k\in \N \cap [0, \log_2 (100n/\eps)+1]$ and $z_0 \in \N^+ \cap  [100/\eps,200/\eps)$, and insert $2^k \cdot z_0$ into $Z$.
	Observe that, every $y\in Y$ incurs a multiplicative error of at most $\eps/100$ after rounding. Hence, $\caS(Z)$ is an $(1-\eps/100)$ approximation of $\caS(Y)$. In particular, $\caS(Z)$ approximates $\caS(Y)$ with additive error at most $(\eps/100) \cdot \Sigma(Y) \le \eps\sigma/100$.
	Combined with previous discussion, this means that $\caS(Z)$ is an $\eps\sigma/50$-additive approximation of $\caS(X)$. 
		
	Then, we process $Z$ using \cref{lem:reduce}, and obtain another set $Z'\subset \N^+$ so that $\caS(Z')=\caS(Z)$, and the multiplicity of any element in $Z'$ is at most $2$. Moreover, by the fourth property of \cref{lem:reduce}, we still have that every $z\in Z'$ can be expressed as $z = 2^k\cdot z_0$ for some non-negative integer $k\le O(\log (n/\eps))$ and integer $z_0 \in \N^+ \cap [100/\eps,200/\eps)$. Now, we can partition $Z'$ into $O(\log (n/\eps))$ groups so that each group contains \emph{distinct} integers from $2^k \cdot \left (\N^+ \cap [100/\eps,200/\eps) \right )$ for some non-negative integer $k\le O(\log (n/\eps))$.
	
	Pick a smaller $\eps' = \Theta \left (\frac{\eps}{ \log(n/\eps)} \right )$ (assuming $\eps/\eps' \in \N^+$). For each group $Z_j'$ mentioned above, we compute a set $A_j \subseteq \N$ that approximates $\caS(Z_j')$ with $\eps'\Sigma(Z_j')/100$ additive error. This can be done  as follows: recall that $Z_j'$ contains distinct integers from $2^k \cdot \left (\N^+ \cap [100/\eps,200/\eps) \right )$; we scale the integers in $Z_j'$ to $ (\eps/\eps')\cdot (\N^+ \cap [100/\eps, 200/\eps\rceil) )$ and then invoke the algorithm for \cref{prob1} which approximates $\caS(Z_j')$ with additive error $|Z_j'| \le \frac{\Sigma(Z'_j)}{(\eps/\eps')\cdot 100/\eps} = \eps' \Sigma(Z_j')/100$ as desired. The total running time for these invocations is (up to $\polylog(n/\eps)$ factors) $\sum_j (|Z_j| + (100/\eps')^{c}) \le \tilde O(n + 1/\eps^c)$.
	
	Now, using the computed $A_j \subseteq \N$ that approximates $\caS(Z_j')$ with $\eps'\Sigma(Z_j')/100$ additive error, we will compute an approximation of $\caS(Z')$ (recall that $Z' = \bigcup_j Z_j'$ is a partition). To do this, we first round every element in every $A_j$ down to integer multiples of $\lceil \eps'\sigma/100\rceil $, and this rounded  $A_j'$ still approximates $\caS(Z_j')$ with additive error at most $\eps'\Sigma(Z_j')/100 + \eps'\sigma/100 \le \eps'\sigma/50$.  Finally, we use FFT to compute the sumset of all these $A'_j$ (there are $O(\log (n/\eps))$ of them), and this will be our approximation of $\caS(Z')$. The accumulated additive error here is at most $O(\log (n/\eps)) \cdot \eps'\sigma/50 \le \eps\sigma/50$, and the running time of these FFTs is $O(\log(n/\eps)) \cdot \tilde O\left (\frac{\sigma}{ \lceil  \eps'\sigma/100\rceil } \right ) \le \tilde O(1/\eps)$.

	We have obtained an $\eps\sigma/50$-additive approximation of $\caS(Z')$. Previously we established $\caS(Z')=\caS(Z)$ and $\caS(Z)$ is an $\eps\sigma/50$-additive approximation of $\caS(X)$, so we have obtained an $\eps\sigma/50$-additive approximation of $\caS(X)$. By \cref{lem:apx}, this is sufficient for solving $(1-\eps)$-approximation \parti on $X$.
	
	The overall running time of this reduction is $\tilde O(n+1/\eps^c)$.
	\end{proof}

\end{document}